\documentclass[landscape,twocolumn,12pt]{article}
\usepackage{graphicx}
\usepackage[utf8x]{inputenc}
\usepackage[margin=0.5in]{geometry}
\usepackage{slashed}
\usepackage{changepage}
\usepackage{sectsty}

\sectionfont{\normalsize}

\usepackage{epsfig,amsmath,amssymb,verbatim,mathrsfs,subfigure}

\def\beq{\begin{eqnarray}}
\def\eeq{\end{eqnarray}}
\def\bea{\begin{eqnarray}}
\def\eea{\end{eqnarray}}

\def\gev{\, {\rm GeV}}

\newcommand{\gsim}{\lower.7ex\hbox{d\;\stackrel{\textstyle>}{\sim}\;$}}
\newcommand{\lsim}{\lower.7ex\hbox{$\;\stackrel{\textstyle<}{\sim}\;$}}

\def\ahad{\Delta\alpha^{(5)}_{had}}
\def\eqx#1{Eq.\,(\ref{#1})}
\def\tablex#1{Table~\ref{#1}}

\def\deltataui{{\delta\tau_i}}

\def\begintable{\begin{table*}}
\def\endtable{\end{table*}}

\newcommand{\as}{\alpha_S}    

\begin{document}

\setlength{\baselineskip}{0.2in}

\begin{titlepage}
\noindent
\begin{flushright}
{\small CERN-PH-TH-2013-285} \\
\end{flushright}
\vspace{-.5cm}

\begin{center}
  \begin{Large}
    \begin{bf}
Study of the 
Standard Model Higgs Boson  \\ \vspace{0.2cm} Partial Widths and Branching Fractions     
   \end{bf}
  \end{Large}
\end{center}
\vspace{0.2cm}
\begin{center}
\begin{large}
Leandro G. Almeida$^a$, Seung J. Lee$^{b,c}$, Stefan Pokorski$^d$, James D.~Wells$^e$ \\
\end{large}
  \vspace{0.3cm}
  \begin{it}

$^{(a)}$LPT, Universit\'e Paris-Sud and CNRS, 91405 Orsay Cedex, France
\vspace{0.2cm}\\
$^{(b)}$Department of Physics, Korea Advanced Institute of Science and Technology, \\335 Gwahak-ro, Yuseong-gu, Daejeon 305-701, Korea
\vspace{0.2cm}\\
$^{(c)}$School of Physics, Korea Institute for Advanced Study, Seoul 130-722, Korea
\vspace{0.2cm}\\
$^{(d)}$Institute of Theoretical Physics, Faculty of Physics, University of Warsaw, Ho\.{z}a 69, Warsaw, Poland
\vspace{0.2cm}\\
$^{(e)}$CERN Theoretical Physics, CH-1211 Geneva 23, Switzerland, and\\
Physics Department, University of Michigan, Ann Arbor MI 48109
 \vspace{0.1cm}
\end{it}

\end{center}

\center{\today}

\begin{abstract}

The discovery of the Higgs boson, with mass known to better than the percent level, enables for the first time precision Higgs boson analyses. Toward this goal, we define an expansion formalism of the Higgs boson partial widths and branching fractions that facilitates such studies. This expansion yields the observables as a perturbative expansion around reference values of Standard Model input observables (quark masses, QCD coupling constant, etc.).  We compute the coefficients of the expansion using state-of-the-art results. We also study the various sources of uncertainties in computing the partial widths and branching fractions more precisely. We  discuss the impact of these results with efforts to discern new physics through precision Higgs boson studies.

\end{abstract}

\vspace{1cm}

\end{titlepage}

\setcounter{page}{2}

\tableofcontents



\section{Introduction\label{sec:intro}}

With the discovery of the Higgs boson~\cite{Higgs Discovery}, particle physics is entering a new era of precision studies of the Higgs sector. The observables are many and include the Higgs boson mass, its total decay width, its spin, its decay branching fractions to Standard Model (SM) particles, its possible decay branching fractions to other exotic final states, and its various production rates at colliders. All of these observables will be studied carefully in time. 

The theory under primary consideration in this article is the Standard Model. The sub-percent-level determination of the Higgs boson mass now enables a complete set of input observables whereby any perturbative high-energy observable involving the Higgs boson can be predicted.

 In this article, our focus is on the careful exposition of the decay partial widths and branching fractions of a SM Higgs boson with mass near $126\gev$.  Our goal is to provide state-of-the-art formulas that can be used in any precision electroweak analysis to investigate compatibility of the data with the SM predictions in these most fundamental and sensitive observables. Other calculations exist in the literature~\footnote{For a basic review, see~\cite{Djouadi:2005gi} and references therein.}, mostly notably from the computer program HDECAY~\cite{Djouadi:1997yw}; however, we wish to provide an independent calculation that includes the latest advances and allows us to vary the renormalization scale in all parts of the computations. This flexibility will be useful in later discussions regarding uncertainties. We also aim to detail the errors that each input into the computation propagates to the final answer for each observable~\cite{Denner:2011mq}. In some cases, these uncertainties are large, and constitute a limitation to how sensitive experimental measurements can be to determining the underlying theory parameters.  Finally, we discuss some implications for physics beyond the SM sensitivities in precision Higgs studies.


\section{Input Observables}

There are an infinite number of SM observables that can be defined, yet any one of them in principle can be computed precisely once a fixed, complete, independent, and finite set of input observables are specified. A convenient set of input observables is
\begin{equation}
{\rm Inputs:}\left\{m_H, M_Z, \ahad, \as (M_Z), m_{f}\right\}, \label{eq:parameters}
\end{equation}
where $m_f$ represents the list of fermion masses of the SM: $m_t, m_b, m_c, m_\tau,m_\mu,$ etc. We are ignoring flavor angles for the purposes of the present discussion.
We can specify $\alpha(M_Z)$ by  $\ahad$ alone. The relation between the two is
\beq
\alpha(M_Z)=\frac{\alpha}{1-\Delta\alpha_{e,\mu,\tau}-\Delta\alpha_t-\ahad},
\label{eq:alpha}
\eeq
where $\alpha$ is the well-known $1/137.036$ and $\Delta\alpha_{e,\mu,\tau}$ and $\Delta\alpha_t$ are perturbatively calculable and known very accurately~\cite{LEPEWWGalpha}. The weak link to a more precise knowledge of $\alpha(M_Z)$ is $\ahad$, which is extracted mostly via dispersion relations from $e^+e^-\to {\rm hadrons}$ data at low energy. Since all the uncertainty of $\alpha(M_Z)$ originates in $\ahad$, it is customary to specify that value as the input, which by \eqx{eq:alpha}, then dictates the value of $\alpha(M_Z)$.  The values of all the input parameters are given in \tablex{tab:param}.

\begintable[t]
\centering
\begin{tabular}{ |c c | c c |}
\hline \hline
$m_H$ & 125.7(4)            & pole mass $m_t$           & 173.5(10)  \\
pole mass $m_c$ & 1.67(7)   & pole mass $m_b$ & 4.78(6)  \\
pole mass $M_Z$ & 91.1535(21) & $G_F$         & 1.1663787(6) $\times 10^{-5}$ \\
pole mass $m_\tau$ & 1.77682(16)         &    $\as (M_Z)$ & 0.1184(7) \\ 
$\alpha (M_Z)$  & ${1/128.96(2)}$     &   $\ahad$         & 0.0275(1) \\
\hline \hline
\end{tabular}
\caption{Reference values for the input observables, see \eqx{eq:parameters}, chosen for computation of the widths and branching ratios of the Higgs boson. Units are in GeV for the masses. All the reference values except for $m_H$~\cite{CMSHiggsMass} and $\alpha (M_Z)$ (or $\ahad$) are given by~\cite{Beringer:1900zz}. $\as (M_Z)$ is taken to be the world average value. As explained in the text, specifying $\alpha(M_Z)$ and $\ahad$, from the Winter 2012 plots of the LEPEWWG~\cite{LEPEWWGalpha}, in this table is redundant but done for convenience of the reader. }
\label{tab:param}
\endtable

The $SU(2)\times U(1)\to U(1)_{EM}$ gauge structure and symmetry breaking make the prediction of  $M_W$ from other input observables an important test of the theory. Nevertheless, for us, $M_W$ shows up in the partial width calculations as a kinematic mass in the propagator of a loop expression ($H\to b\bar b$ loops, etc.)\ or as the final state mass in a phase space computation ($H\to WW^*$). Since it very directly appears in these computations, one might be tempted to  choose it as an input parameter to the calculations of precision Higgs observables. This is legitimate and acceptable in principle. One can exchange, for example, $m_t$ for $M_W$ as an input. However, several complications arise. The choice of $M_W$ as an input may simplify the computation in some ways, but makes it more complicated in other ways (e.g., utilizing self-consistent top mass). More importantly, there is a risk that by doing so, one can choose incompatible sets of input parameters for predictions of different sets of observables. For example, in computing precision $Z$-decay and LEP2 observables, one might choose the standard set of inputs that does not include $M_W$, whereas for Higgs sector observables, one might choose a set of inputs that includes $M_W$.  Making a comparison between $B(Z\to b\bar b)$ and $B(H\to b\bar b)$, for example, when testing the SM becomes impossible unless an equivalence dictionary between the two sets is clearly specified, and self-consistent, equivalent sets of inputs are chosen. For this reason, we specify one set of input observables for all computations, and that set is the one where $M_W$ is an output.

Now that we have established  our convention that $M_W$ is an output observable,  when the $W$ mass appears in formulas below, we should view it as a short-hand notation for the full computation of the $W$ mass within the theory in terms of our agreed-upon inputs. In the SM this substitution is
\bea
M_W & \stackrel{SM~}\longrightarrow &(80.368\gev)\left(1+1.42\,\delta M_Z+0.21\,\delta G_F-0.43\, \delta\alpha\right. \nonumber \\
& & ~~~~\left. +0.013\,\delta M_t
-0.0011\,\delta\alpha_S -0.00075\,\delta M_H\right).
\label{eq:MW replacement}
\eea
This formula is obtained by expanding results in~\cite{Ferroglia:2002rg}, which we have independently checked.  Numerical evaluation was done using the reference values of the input parameters given in \tablex{tab:param}.  The definition of $\delta \tau$ is $\delta \tau\equiv (\tau-\tau_{ref})/\tau_{ref}$.


One ultimate goal of this work is to survey state-of-the art calculations in order to test the SM. The proper way to test any theory is to compute all the observables and subject it to a global $\chi^2$ likelihood test, where 
\beq
\chi^2=\sum_i \left( \frac{{\cal O}^{th}-{\cal O}^{expt}}{\Delta{\cal O}^{expt}_i}\right)^2,
\eeq
one should take all the correlations amongst observables into account as well~\cite{Espinosa:2012ir}. Upon computing the $\chi^2$ it is then possible to ask statistical inference questions to the value. For example, is the $\chi^2$ per degree of freedom indicating the theory is compatible with the data at some confidence level? In such a procedure, it makes no difference what independent input parameters one uses: there is an infinite set of possibilities that are equally good and the answer is the same to any well-defined question regarding confidence in a theory or range of values predicted for an observable given the data, etc.  

The results that we present will also enable a very quick determination of our present abilities to determine from the measurement couplings of SM particles to the Higgs boson. Each of the input observables has a number of uncertainties associated with it, and when these errors propagate, there will be uncertainties for the predictions of the partial widths and branching fraction observables.  
At the moment, the predicted uncertainties (a few percent or less) are much smaller than the current measured uncertainties (tens of percent), but in the future this limiting theory precision will become important as experiments improve. We note that in the current experimental situation the prediction uncertainties are nearly the same had we chosen $M_W$ rather than $m_t$ as an input. Indeed it is somewhat accidental that the target observables have nearly the same small prediction uncertainty for either choice.

In the following sections, we will present the computations for each of the important decay-mode partial widths of the SM Higgs boson. The results are presented here in order to show the origin of our subsequently derived  expansions of these partial widths and branching fractions in terms of small deviations away from measured reference values of the input observables.

\section{Higgs Boson Partial Widths}

In this section, we describe the procedures by which we compute the partial widths of the Higgs boson decays. In each case they are taken from state-of-the-art computations within the literature. It is intended that the reader can reproduce all of our results by following the instructions we give below.

\medskip
\noindent
{\it Higgs decays to $WW^*$ and $ZZ^*$}
\medskip

The interaction between the Higgs and the electroweak vector bosons can be best probed through its direct decay into  vector bosons. The mass of Higgs boson at 126 GeV excludes the decay into two on-shell electroweak vector bosons and leaving the following decays, $H\to V^{(*)} V^*$. The width for a Higgs boson decaying into $V^*V$ is given at lowest order in~\cite{Keung:1984hn}. 

The vector bosons further decay into fermions, and there are interferences between the intermediate off-shell vector bosons as a consequence. This result is known at ${\cal{O}}(\as)$ and at ${\cal{O}}(\alpha)$ and its calculation is described in \cite{Bredenstein:2006rh}. The results were implemented in a Monte-Carlo generator, {\it Prophecy4f} \cite{Bredenstein:2007ec}, which is what we use to compute the $WW^*$ and $ZZ^*$ partial width values.

\medskip\noindent
{\it Higgs decays to $\gamma\gamma$ and $Z\gamma$}
\medskip

The higher order contributions to  $\Gamma(H\to\gamma \gamma)$ are known to NNLO ${\cal{O}}(\as^2)$ in QCD \cite{Maierhofer:2012vv}, and at NLO in purely EW Corrections \cite{Actis:2008ug}. We parametrize the results as, 
\begin{eqnarray}
\Gamma(H\to \gamma \gamma) & = &  \Gamma^{\gamma \gamma}_{LO} + \frac{\alpha}{\pi} \Gamma^{\gamma \gamma}_{NLO-EW} +\left(\frac{\as}{\pi}\right) \Gamma^{\gamma \gamma}_{NLO-QCD}\nonumber \\
& & ~~~+\left(\frac{\as}{\pi}\right)^2 \Gamma^{\gamma \gamma}_{NNLO-QCD}.
\end{eqnarray}
where  $\alpha=\alpha_{QED}(m_H^2)$ at one-loop, and $\as=\as(m_H^2)$ at 3-loop as provided by RunDec~\cite{Chetyrkin:2000yt}. The results for $\Gamma_{NNLO-QCD}$ are obtained from \cite{Maierhofer:2012vv}, and for consistency, we also use its results for $\Gamma_{NLO-QCD}$. For both orders we use the expansion in $x_t=m^2_H/(4m_t^2)$ to ${\cal{O}}(x_t^5)$. 
For $\Gamma_{NLO-EW}$ we interpolate the results of \cite{Actis:2008ug} to the same order. 

For the prediction of $\Gamma(H\to Z \gamma)$ we use the results of~\cite{Spira:1991tj}, which give the contributions to lowest order with an additional contribution from QCD involving top quark loops. 
The result is parametrized below, 
\begin{equation}
 \Gamma (H\to Z \gamma)  =  \Gamma^{Z \gamma}_{LO} + \frac{\as}{\pi} \Gamma^{Z \gamma}_{\rm{aprox-NLO}}.
\end{equation}  
Here $\Gamma^{Z \gamma}_{\rm{aprox-NLO}}$ is the additional contribution from QCD to the top quark loop. This is achieved by shifting the top amplitude in the lowest order contribution~\cite{Spira:1991tj}.

\medskip\noindent {\it Higgs decays to gluons}\medskip

Similarly, the partial width of Higgs decaying into gluons is given at NNLO, ${\cal{O}}(\as^2)$ in the full SM theory \cite{Schreck:2007um}. While the result is also known at NNNLO in the effective theory~\cite{Chetyrkin:1997sg,Steinhauser:2002rq}, resulting from integrating out the top quark, we use only the results from full SM computation.

For the electroweak corrections we make use of the numerical results of~\cite{Actis:2008ts}, and extrapolate them to ${\cal{O}}(x_t^4)$, where $x_t=m_H^2/(4 m^2_t)$.  We use the 3-loop result for $\as$, running to the proper scale choice. The scales are chosen to be $m_H$, with the exception of the electroweak corrections, whose scale dependences are not provided, but were indicated to be small in~\cite{Actis:2008ts}. 

\medskip\noindent{\it Higgs decays to quarks}\medskip

The dominant decay for the Higgs is directly into $b \bar{b}$. For its partial width
we use the results of~\cite{Baikov:2005rw}, which provide the non-power-suppressed corrections to  
${\cal{O}}(\as^4)$. We obtain ${\cal{O}}(G_F m_t^4)$ corrections from~\cite{Butenschoen:2007hz}.
Higher-order logarithmic corrections are absorbed into the 
running quark masses. All masses are evolved using functions 
obtained from RunDec~\cite{Chetyrkin:2000yt} to the appropriate loop order. In the case of the $H\to b\bar{b}$ partial width, we need to evolve the $\overline{MS}$ mass, $m_b$, to 3 loops, 
given the accuracy of the calculation. In the case of $c\bar{c}$, we make use of the electroweak corrections found in \cite{Butenschoen:2007hz},
while keeping the same order in QCD as $b\bar{b}$.
The scale dependence to order ${\cal{O}}(\as^3)$ is given in \cite{Chetyrkin:1996sr,Chetyrkin:1997vj} for the diagonal correlators.  We make use of the result at ${\cal{O}}(\as^4)$ at $s=m_h^2$ and $n_f=5$ from \cite{Baikov:2005rw}, and with the renormalization group equations, extend the scale dependence to ${\cal{O}}(\as^4)$.
The one-loop pure electroweak contributions were obtained by \cite{Hollik:1992,Kniehl:1992lo}. We use the full analytical result for its dependence on all on-shell quark masses (with $m_{u,d,s}=0$) and lepton masses ($m_e=0$). The $W$ mass is determined as described above, and we subtract the leading contribution in $G_F m_t^2$ to avoid double counting from the contributions mentioned above.  

\medskip\noindent{\it Higgs to leptons}\medskip

For the partial decay width into two leptons we make use of the next-to-leading-order QCD corrections up to ${\cal{O}}(\alpha_s^2 G_F m_t^2)$ and 2-loop electroweak corrections in~\cite{Kniehl:1995br,Butenschoen:2007hz}.  

\section{Expansion of Partial Widths and Uncertainties\label{sec:partial}}

Now that we have the full expressions for the partial widths of Higgs decays we are in position to Taylor expand these equations around the input observables.  This expansion is made possible by the fact that with the discovery of the Higgs boson, and knowledge of its mass, all input observables  are now known to good enough accuracy to render an expansion of this nature useful and accurate. 

We represent the partial width expansion by
\begin{equation}\label{eq:partialwidthdefition}
	\Gamma_{H\to X} = \Gamma_X^{(\rm{ref})} \left ( 1 + \sum_i a_{\tau_i,X} \overline{\deltataui}\right)
\end{equation}
where 
\begin{equation}
	\overline{\deltataui}= \frac{\tau_i- \tau_{i,ref}}{\tau_{i,ref}},
\end{equation}
and $\tau_i$ are the input observables (Eq.~\ref{eq:parameters}) for the calculation. The total width is the sum of all the partial widths and for convenience we present dedicated expansion parameters for that as well:
\begin{equation}
\Gamma_{\textrm{tot}} =\sum_{X}\Gamma_{H\to X} = \Gamma_{\textrm{tot}}^{(\rm{ref})} \left ( 1 + \sum_i a_{\tau_i,\textrm{tot}} \overline{\deltataui}\right).
\end{equation}

For many of our parameters and observables we would like to know the relative uncertainty due to variations in the input parameters or variations of scale.  The ``percent relative uncertainty" $P_Q$ of a parameter or observable $Q$ from its central reference value $Q_0$ is defined to be
\beq
Q=Q_0\, (1+0.01\,P_Q).
\label{eq:percenterror}
\eeq
If the errors are asymmetric then $P^+_Q$ designates the positive percent relative error that increases the absolute value of $Q$, and $P^-_Q$ designates the negative percent relative error that decreases the absolute value of $Q$.  If the positive and negative errors are symmetric then we can combine and label it as $P^\pm_Q$.

In the computation we need to know the quark masses at the scale of the Higgs boson mass. The quark masses for bottom and charm are evolved to $\mu = m_H$ using renormalization group techniques. For consistency, and to avoid confusion, we use the program RunDec~\cite{Chetyrkin:2000yt}, which provides the $\overline{\rm{MS}}$ evolution up to ${\cal{O}}(\as^4)$. The values at $\mu =m_H$ are provided in \tablex{tab:runmass} for easy comparisons.
	\begintable[t]
		\centering
	\begin{tabular}{| c | c | c | }
	\hline \hline 
	quark & at $\mu=m_H\, (m_H/2,\,2m_H)$ & $P_{m}(\Delta m)$ \\
	\hline 
	$m_c (\mu)$ & 0.576 (0.612, 0.546) GeV & 7.53\% \\ 
	$m_b (\mu)$ & 2.68 (2.84, 2.54) GeV & 1.62\% \\ 
	$m_t (\mu)$ & 167 (177,158) GeV & 0.63\% \\ 
	\hline \hline
	\end{tabular}
	\caption{Running $\overline{\rm MS}$ masses for the heavy quarks at 3-loops at the scale $\mu=m_H$, $m_H/2$ and $2m_H$ from program RunDec~\cite{Chetyrkin:2000yt}, which is used for the Higgs decaying into quarks. Pole-mass inputs are taken from \tablex{tab:param}. The parametric uncertainty on the running mass at $\mu=m_H$ from $1\sigma$ uncertainty ($\sigma_m$) in the pole mass is defined to be $P_m(\Delta m)=\{ m_+(m_H)+m_-(m_H)\}/\{2m(m_H)\}$, where $m_\pm(m_H)$ is computed using $m_{\rm pole}=m_{\rm ref}\pm \sigma_m$. The scale dependence of the running mass is cancelled in higher order loop calculations, as can be seen later for scale-dependence uncertainties.}
	\label{tab:runmass}
	\endtable

The partial width expansions are given in \tablex{tab:analresults1}, where the reference values $\Gamma^{(ref)}_0$ and expansion coefficients  of \eqx{eq:partialwidthdefition} are computed using the central values of the input observables of \tablex{tab:param}.  \tablex{tab:theoreticaluncertainties} gives the estimated parametric and scale-dependence uncertainties on the partial width values of Table~\ref{tab:analresults1}. 
The uncertainties are expressed as ``percent relative uncertainties" according to the definition in \eqx{eq:percenterror}. The meaning of ``$P^\pm_\Gamma({\rm par.add.})$" is that all input parameters have been allowed to range over their $1\sigma$ errors and the maximum percent relative errors are recorded. The meaning of ``$P^\pm_\Gamma({\rm par.quad.})$" is that the uncertainties of each parameter are added in Gaussian quadrature. In other words, $P_{\Gamma_i}^\pm{\rm (par.quad.)}=100\,\Delta\Gamma_i/\Gamma_i$, where 
\beq
(\Delta \Gamma_i)^2=\left( \frac{\partial\Gamma_i}{\partial m_t}\right)^2 (\Delta m_t)^2
+\left( \frac{\partial\Gamma_i}{\partial \alpha_s}\right)^2 (\Delta \alpha_s)^2+\cdots.
\eeq

The uncertainties in varying the scale parameter $\mu$ in the calculation, attempts to capture the uncertainty in not knowing higher order corrections. A full calculation at all orders would give a result that does not depend on $\mu$ but a finite-order calculation does, and the uncertainty of dropping the higher order calculations are assumed to be approximated reasonably well by noting how much the result changes by varying $\mu$ by a factor of two upward and downward: $m_H/2<\mu<2m_H$. The meaning of ``$P^\pm_\Gamma(\mu)$" in \ \tablex{tab:theoreticaluncertainties} concerns the relative percent uncertainties associated with this scale dependence algorithm.

\begintable[t]
\begin{center}
                \footnotesize
	\begin{tabular}{|c|c|c|c|c|c|c|c|c|c|c|c|}
		\hline\hline
 &   $\Gamma_X^{\rm(Ref)}/{\rm GeV}$ & $a_{m_t,X}$ &  $a_{m_H,X}$ &  $a_{\alpha ( M_Z),X}$ & $a_{\as ( M_Z),X}$ &  $a_{m_b,X}$ &  $a_{M_Z,X}$ &  $a_{m_c,X} $ & $a_{m_\tau,X}$ & $a_{G_F,X}$\\
	  \hline 
 \textrm{total} & 3.96$\times 10^{-3}$ & -3.48$\times 10^{-2}$ & 4.53 & 8.77$\times 10^{-1}$ & -1.35 & 1.4 & -3.49 & 9.05$\times 10^{-2}$ & 1.3$\times 10^{-1}$ & 8.43$\times 10^{-1}$ \\
$gg$ & 3.57$\times 10^{-4}$ & -1.62$\times 10^{-1}$ & 2.89 & 0. & 2.49 & -7.1$\times 10^{-2}$ & 3.77$\times 10^{-1}$ & 0. & 0. & 1. \\
$\gamma\gamma$ & 1.08$\times 10^{-5}$ & -2.73$\times 10^{-2}$ & 4.32 & 2.56 & 1.8$\times 10^{-2}$ & 9.01$\times 10^{-3}$ & -1.85 & 0. & 0. & 7.24$\times 10^{-1}$ \\
$b\bar{b}$ & 2.17$\times 10^{-3}$ & 8.11$\times 10^{-3}$ & 8.09$\times 10^{-1}$ & 3.76$\times 10^{-2}$ & -2.46 & 2.57 & -4.75$\times 10^{-1}$ & 0. & 0. & 9.53$\times 10^{-1}$ \\
 $c\bar{c}$ & 9.99$\times 10^{-5}$ & -4.55$\times 10^{-2}$ & 7.99$\times 10^{-1}$ & 1.02$\times 10^{-2}$ & -9.17 & 0. & -1.41 & 3.59 & 0. & 9.7$\times 10^{-1}$ \\
$\tau^+\tau^-$ & 2.58$\times 10^{-4}$ & 4.74$\times 10^{-2}$ & 9.95$\times 10^{-1}$ & -2.09$\times 10^{-2}$ & -2.15$\times 10^{-3}$ & 0. & -1.61$\times 10^{-2}$ & 0. & 2.01 & 1.02 \\
$WW^*$ & 9.43$\times 10^{-4}$ & -1.13$\times 10^{-1}$ & 1.37$\times 10^1$ & 3.66 & 9.04$\times 10^{-3}$ & 0. & -1.21$\times 10^1$ & 0. & 0. & 2.49$\times 10^{-1}$ \\
$ZZ^*$ & 1.17$\times 10^{-4}$ & 2.28$\times 10^{-2}$ & 1.53$\times 10^1$ & -7.37$\times 10^{-1}$ & -1.82$\times 10^{-3}$ & 0. & -1.12$\times 10^1$ & 0. & 0. & 2.53 \\
$ Z\gamma$ & 6.88$\times 10^{-6}$ & -1.54$\times 10^{-2}$ & 1.11$\times 10^1$ & 8.46$\times 10^{-1}$ & 0. & -9.76$\times 10^{-3}$ & -4.82 & 0. & 0. & 2.62 \\
 $\mu^+\mu^-$ & 8.93$\times 10^{-7}$ & 4.84$\times 10^{-2}$ & 9.92$\times 10^{-1}$ & -4.31$\times 10^{-2}$ & -2.2$\times 10^{-3}$ & 0. & -1.62$\times 10^{-2}$ & 0. & 0. & 1.02 \\

	\hline \hline                       
	\end{tabular}
	\caption{Reference values for the partial widths at the central values of the parameters given in 
	         \tablex{tab:param} along with values for $a_{\tau_i,X}$ as
  			 defined by \eqx{eq:partialwidthdefition}. $VV^*$ partial decay widths are calculated by {\it Prophecy4f}.}.
	 \label{tab:analresults1}
	 \end{center}
	\endtable

\begintable[t]
\begin{center}
	\begin{tabular}{|c|c|c|c|}
		\hline\hline
 &  $P_\Gamma^\pm(\text{par.add.})$ & $P_\Gamma^\pm(\text{par.quad.})$ & $(P_\Gamma^+,\, P_\Gamma^-)(\mu)$ \\
	  \hline 
\textrm{total} & 4.41 (3.33) & 2.43 (2.00) & (0.06,0.09) \\
$gg$ &  2.57 (1.88) & 1.74 (1.50) & (0.01,0.04) \\
$\gamma\gamma$ & 1.46 (0.43) & 1.38 (0.35) & (1.31,0.60) \\
$b\bar{b}$ & 4.94 (4.75) & 3.54 (3.53) & (0.31,0.02) \\
$c\bar{c}$ & 20.75 (20.56) & 15.99 (15.99) & (0.43,0.32) \\
$\tau^+\tau^-$  & 0.36 (0.13) & 0.32 (0.09) & (0.01,0.01) \\
$WW^*$ & 4.43 (1.17) & 4.97 (1.25) & (0.25,0.31) \\
$ZZ^*$ & 4.90 (1.25) & 4.42 (1.11) & (0.,0.) \\
$Z\gamma$ & 3.57 (0.93) & 3.52 (0.88) & (0.56,0.23) \\
$\mu^+\mu^-$  & 0.35 (0.11) & 0.32 (0.08) & (0.03,0.03) \\
  	\hline \hline                       
	\end{tabular}
	\caption{This table gives the estimates for percent relative uncertainty on the partial widths from parametric and scale-dependence uncertainties. Parametric uncertainties arise from incomplete knowledge of the input observables for the calculation (i.e., errors on $m_c$, $\alpha_s$, etc.). For parametric uncertainties, we put an additional number in parentheses, which is the value it would have if the Higgs mass uncertainty were 0.1 GeV (instead of 0.4 GeV).
	Scale-dependence uncertainties are indicative of not knowing the higher order terms in a perturbative expansion of the observable. These uncertainties are estimated by varying $\mu$ from $m_H/2$ to $2m_H$.  More details on the precise meaning of the entries of this table are found in the text of sec.~\ref{sec:partial}. Errors below $0.01\%$ are represented in this table as 0.}
	 \label{tab:theoreticaluncertainties}
\end{center}
	\endtable

\section{Expansion of Branching Fractions and Uncertainties}

In the previous section we derived the expansion of the partial widths in terms of small deviations of the input observables from their reference values, and we determined the uncertainties of the partial widths due to input observable uncertainties (parameter uncertainties) and scale-dependence uncertainties. The same type of expansion can be done for branching fractions, and ratios of branching fractions. To begin with, the expansion for the branching ratios are 
\begin{equation}\label{eq:branchingfractiondefition}
	\rm{B}({\it{H}\to \rm{X}}) = \rm{B}(X)^{(\rm{ref})} \left ( 1 + \sum_i b_{\tau_i,X} \overline{\deltataui}\right),
\end{equation}
where $\tau_i$ represents the same parameters as \eqx{eq:parameters}. 
Expansion parameters $b_{\tau_i,X}$ are related to $a_{\tau_i,X}$ by
\begin{equation}\label{eq:bbya}
	b_{\tau_i,X} =a_{\tau_i,X}-a_{\tau_i,tot}.
\end{equation}
Using the
reference parameters from \tablex{tab:param} we display the results of the reference branching ratios and their expansion coefficients in \tablex{tab:BRfractions1}.

 \begintable[t]  %
 \begin{center}
	\footnotesize
	\begin{tabular}{|c|c|c|c|c|c|c|c|c|c|c|}
		\hline \hline
&  $\rm{B}(X)^{(\rm{Ref})}$ &  $b_{m_t}$ &  $b_{m_H}$ &  $b_{\alpha( M_Z)}$ &  $b_{\as( M_Z)}$ &  $b_{m_b}$ &  $b_{M_Z}$ &  $b_{m_c} $ & $b_{m_\tau}$ & $b_{G_F}$ \\
	   \hline 	
$gg$ & 9.03$\times 10^{-2}$ & -1.27$\times 10^{-1}$ & -1.64 & -8.77$\times 10^{-1}$ & 3.84 & -1.47 & 3.87 & -9.05$\times 10^{-2}$ & -1.30$\times 10^{-1}$ & 1.57$\times 10^{-1}$ \\
$\gamma\gamma$ & 2.73$\times 10^{-3}$ & 7.46$\times 10^{-3}$ & -2.1$\times 10^{-1}$ & 1.68 & 1.37 & -1.39 & 1.64 & -9.05$\times 10^{-2}$ & -1.30$\times 10^{-1}$ & -1.19$\times 10^{-1}$ \\
$b\bar{b}$ & 5.47$\times 10^{-1}$ & 4.29$\times 10^{-2}$ & -3.72 & -8.40$\times 10^{-1}$ & -1.11 & 1.17 & 3.02 & -9.05$\times 10^{-2}$ & -1.30$\times 10^{-1}$ & 1.10$\times 10^{-1}$ \\
$c\bar{c}$  & 2.52$\times 10^{-2}$ & -1.07$\times 10^{-2}$ & -3.73 & -8.67$\times 10^{-1}$ & -7.82 & -1.40 & 2.08 & 3.50 & -1.30$\times 10^{-1}$ & 1.26$\times 10^{-1}$ \\
$\tau^+\tau^-$ & 6.51$\times 10^{-2}$ & 8.22$\times 10^{-2}$ & -3.53 & -8.98$\times 10^{-1}$ & 1.35& -1.40 & 3.48 & -9.05$\times 10^{-2}$ & 1.87 & 1.72$\times 10^{-1}$ \\
$WW^*$ & 2.38$\times 10^{-1}$ & -7.87$\times 10^{-2}$ & 9.14 & 2.78 & 1.36 & -1.40 & -8.63 & -9.05$\times 10^{-2}$ & -1.30$\times 10^{-1}$ & -5.94$\times 10^{-1}$ \\
$ZZ^*$ & 2.96$\times 10^{-2}$ & 5.76$\times 10^{-2}$ & 1.08$\times 10^1$ & -1.61 & 1.35 & -1.40 & -7.69 & -9.05$\times 10^{-2}$ & -1.30$\times 10^{-1}$ & 1.69 \\
$Z\gamma$ & 1.74$\times 10^{-3}$ & 1.94$\times 10^{-2}$ & 6.53 & -3.12$\times 10^{-2}$ & 1.35 & -1.41 & -1.32 & -9.05$\times 10^{-2}$ & -1.30$\times 10^{-1}$ & 1.78 \\
$\mu^+\mu^-$ & 2.25$\times 10^{-4}$ & 8.32$\times 10^{-2}$ & -3.53 & -9.20$\times 10^{-1}$ & 1.35 & -1.40 & 3.48 & -9.05$\times 10^{-2}$ & -1.30$\times 10^{-1}$ & 1.73$\times 10^{-1}$ \\
 	  \hline \hline  
	\end{tabular}
	\caption{The reference value and expansion coefficients for Higgs boson decay branching fractions according to \eqx{eq:branchingfractiondefition}. The input parameters for this computation are from \tablex{tab:param}.
	 $VV^*$ partial decay widths are calculated by {\it Prophecy4f}.}
	\label{tab:BRfractions1}
\end{center}
\endtable

 \begintable[t]  %
 \begin{center}
	\footnotesize
	\begin{tabular}{|c|c|c|c|c|c|c|c|c|c|}
		\hline \hline
& $\Delta_{m_t}$ & $\Delta_{m_H}$ & $\Delta_{\alpha(M_Z)}$ & $\Delta_{\as(M_Z)}$ & $\Delta_{m_b}$ & $\Delta_{M_Z}$ & $\Delta_{m_c}$ & $\Delta_{m_\tau}$ & $\Delta_{G_F}$ \\		
	   \hline 	
$gg$ & 0.07 & 0.52 (0.13) & 0.01 & 2.27 & 1.84 & 0.01 & 0.38 & - & - \\	   
$\gamma\gamma$ & - & 0.07 ( 0.02 ) & 0.03 & 0.81 & 1.74 & - & 0.33 & - & - \\
$b\bar{b}$ & 0.02 & 1.18 (0.30) & 0.01 & 0.66 & 1.47 & 0.01 & 0.38 & - & - \\
$c\bar{c}$ & 0.01 & 1.19 (0.30) & 0.01 & 4.62 & 1.75 & - & 14.66 & - & - \\
$\tau^+\tau^-$ & 0.05 & 1.12 (0.28) & 0.01 & 0.80 & 1.75 & 0.01 & 0.38 & 0.02 & - \\
$WW^*$ & 0.05 & 2.91 (0.73) & 0.04 & 0.80 & 1.75 &0.02 & 0.38 & - & - \\
$ZZ^*$ & 0.03 & 3.43 (0.86) & 0.02 & 0.80 & 1.75 & 0.02 & 0.38 & - & - \\
$Z\gamma$ & 0.01 & 2.08 (0.52) & - & 0.80 & 1.76 & - & 0.38 & - & - \\
$\mu^+\mu^-$ & 0.05 & 1.12 (0.28) & 0.01 & 0.80 & 1.75 & 0.01 & 0.38 & - & - \\
 	  \hline \hline  
	\end{tabular}
	\caption{Table of percentage uncertainties of branching fractions due to uncertainties in each of the input observables, as calculated by eq.~\ref{eq:percentuncertainty}. The input parameters for this computation are from \tablex{tab:param}. In addition we also compute the branching ratio uncertainties due to $\Delta m_h = 0.1$ GeV, the expected uncertainty after LHC run. These values are in parenthesis in the $\Delta_{m_H}$ column. Percentages less than $0.1\%$ are listed as $-$.}
	\label{tab:newBRerror}
\end{center}
\endtable

 
\begintable[t]
 \begin{center}
	\begin{tabular}{|c|c|c|c|}
		\hline\hline
 &  $P_\text{BR}^\pm(\text{par.-add.})$ & $P_\text{BR}^\pm(\text{par.-quad.})$ & $(P_\text{BR}^+,\, P_\text{BR}^-)(\mu)$ \\
	  \hline 
$gg$ & 5.11 (4.72) & 2.99 (2.95) & (0.01,1.22) \\
$\gamma\gamma$  & 3.03 (2.98) & 1.96 (1.96) & (1.80,1.81) \\
$b\bar{b}$ & 3.73 (2.85) & 2.03 (1.68) & (0.24,0.00) \\
$c\bar{c}$ & 22.25 (21.36) & 15.52 (15.48) & (0.52,0.38) \\
$\tau^+\tau^-$  & 4.14 (3.30) & 2.26 (1.98) & (0.08,0.05) \\
$WW^*$ & 5.95 (3.77) & 3.51 (2.10) & (0.09,0.06) \\
$ZZ^*$ & 6.43 (3.86) & 3.95 (2.14) & (0.09,0.06) \\
$Z\gamma$ & 5.04 (3.48) & 2.87 (2.04) & (0.83,0.78) \\
$\mu^+\mu^-$  & 4.12 (3.28) & 2.26 (1.98) & (0.07,0.04) \\
 	\hline \hline                       
	\end{tabular}
	\caption{This table gives the estimates for  theory error (percent relative uncertainty) of the branching fractions due to  parametric uncertainties and due to scale-dependent uncertainties from varying $m_H/2 \le \mu \le 2m_H$. Errors below 0.01\% are reported in the table as 0. For parametric uncertainties, we put an additional number in parentheses, which is the value it would have if the Higgs mass uncertainty were 0.1 GeV (instead of 0.4 GeV).}
	 \label{tab:BRfractionsUncertainties}
\end{center}
	\endtable

The table of expansion coefficients enables us to compute the uncertainty in a final state branching ratio due to each input parameter.  The percent uncertainty $\Delta^X_i$ on branching fraction $B(X)$ due to input parameter $\tau_i$ is
\beq
\Delta^X_i=(100\%)\times |b_{\tau_i,X}| \, \frac{\Delta\tau_i}{\tau_i^{ref}}
\label{eq:percentuncertainty}
\eeq
where $\Delta\tau_i$ are the current experimental uncertainties in input parameter $\tau_i$. For example, the percentage uncertainty in the $H\to gg$ branching fraction is 
\beq
\Delta^{gg}_b=(100\%)(1.467) \frac{0.06\gev}{4.78\gev}=1.84\%.
\eeq
Each of these calculations have been done and are presented in Table~\ref{tab:newBRerror}. 
We see most clearly in this table that the uncertainty in the $b$-quark mass input observable constitutes the largest uncertainty in the branching ratio computations. The large uncertainty of the charm quark mass is the decisive contributor to $H\to c\bar c$ uncertainty as well.

\bigskip
\noindent
{\it Ratios of branching ratios}
\medskip

Experimental observables at colliders are a combination of cross-section times branching fraction, $\sigma B$. Although this combination $\sigma B$ can be often measured to very high accuracy, the extraction of the branching fractions are fraught with experimental and computational complexity in several ways. First, the parton distribution functions are not known with high enough precision to perform a calculation that would match the precision with which the observable ultimately will be measured. Second, there are additional theory uncertainties in the cross-section and the definition of the overall observable that make difficult the clean comparison between theory and experiment.  For this reason, it is often useful~\cite{Djouadi:2013qya} to measure ratio of observables $(\sigma B_1)/(\sigma B_2)\simeq B_1/B_2$, where the uncertainties in the production cross-section largely drop out. It is beyond the purpose and scope of this paper to detail this process, but what we can do now is give accurate computations of the uncertainties of the ratios of branching fractions.

As with the partial widths and the branching fractions themselves, it is useful to expand  the ratio of the branching fractions in the following form
\begin{equation}\label{eq:ratiobranchingfractiondefition}
	\frac{\rm{B}({\it{H}\to \rm{X}})}{\rm{B}({\it{H}\to \rm{Y}})} = \frac{\rm{B}(X)^{(\rm{ref})}}{\rm{B}(Y)^{(\rm{ref})}} \left ( 1 + \sum_i r_{\tau_i,X,Y} \overline{\deltataui}\right),
\end{equation}
where $\tau_i$ represent the same parameters as \eqx{eq:parameters}.
The expansion parameters $r_{\tau_i,X,Y}$ is related to $a_{\tau_i,X}$ by
\begin{equation}\label{eq:bbyatwo}
	r_{\tau_i,X,Y} =a_{\tau_i,X}-a_{\tau_i,Y}.
\end{equation}
Using the
reference parameters from Table \ref{tab:param}, we display the results of the reference ratio of the branching ratios and their deviations in \tablex{tab:BRratio}. \tablex{tab:BRratioUncertainties} presents the uncertainties of the predictions for these observables. We see that typically there is a few percent uncertainty in predicting the ratios of branching ratios, and as emphasized above these may be the cleanest observables the LHC experiment will present us for some time.

\begintable[t]
\begin{center}
		\footnotesize
\begin{adjustwidth}{-0.0cm}{}	
	\begin{tabular}{|c|c|c|c|c|c|c|c|c|c|c|c|c|}
		\hline\hline
{} & $B(X)/B(Y)_{Ref}$ & $r_{m_t}$ & $r_{m_H}$ & $r_{\alpha(M_Z)}$ & $r_{\as(M_Z)}$ & $r_{m_b}$ & $r_{M_Z}$ & $r_{m_c}$ & $r_{m_\tau}$  & $r_{G_F}$ \\
	   \hline 	
$\gamma \gamma /WW^*$ & 1.15$\times 10^{-2}$ & 8.62$\times 10^{-2}$ & -9.35 & $-1.10$ & 8.95$\times 10^{-3}$ & 9.01$\times 10^{-3}$ & 1.03$\times 10^1$ & 0. & 0. & 4.75$\times 10^{-1}$ \\
$b\bar{b}/c\bar{c}$ & 2.17$\times 10^1$ & 5.36$\times 10^{-2}$ & 9.6$\times 10^{-3}$ & 2.74$\times 10^{-2}$ & 6.71 & 2.57 & 9.34$\times 10^{-1}$ & -3.59 & 0. & -1.66$\times 10^{-2}$ \\
$\tau^+ \tau^- /\mu^+ \mu^- $ & 2.89$\times 10^2$ & -1.03$\times 10^{-3}$ & 2.55$\times 10^{-3}$ & 2.22$\times 10^{-2}$ & 4.65$\times 10^{-5}$ & 0. & 1.09$\times 10^{-4}$ & 0. & 2.01 & -3.39$\times 10^{-4}$ \\
$c\bar{c}/\mu^+ \mu^-$  & 1.12$\times 10^2$ & -9.39$\times 10^{-2}$ & -1.93$\times 10^{-1}$ & 5.33$\times 10^{-2}$ & -9.17 & 0. & -1.39 & 3.59 & 0. & -4.64$\times 10^{-2}$ \\
$WW^*/ZZ^*$ & 8.05 & -1.36$\times 10^{-1}$ & -1.63 & 4.40 & 1.09$\times 10^{-2}$ & 0. & -9.38$\times 10^{-1}$ & 0. & 0. & -2.28 \\
$\gamma\gamma/ZZ^*$ & 9.22$\times 10^{-2}$ & -5.02$\times 10^{-2}$ & -1.10$\times 10^1$ & 3.30 & 1.98$\times 10^{-2}$ & 9.01$\times 10^{-3}$ & 9.33 & 0. & 0. & -1.81 \\
$b\bar{b}/ZZ^*$ & 1.85$\times 10^1$ & -1.47$\times 10^{-2}$ & -1.45$\times 10^1$ & 7.74$\times 10^{-1}$ & -2.46 & 2.57 & 1.07$\times 10^1$ & 0. & 0. & -1.58 \\
$\tau^+\tau^-/ZZ^*$ & 2.2 & 2.46$\times 10^{-2}$ & -1.43$\times 10^1$ & 7.16$\times 10^{-1}$ & -3.29$\times 10^{-4}$ & 0. & 1.12$\times 10^1$ & 0. & 2.01 & -1.52 \\
$Z\gamma /ZZ^*$ & 5.87$\times 10^{-2}$ & -3.82$\times 10^{-2}$ & -4.23 & 1.58 & 1.82$\times 10^{-3}$ & -9.76$\times 10^{-3}$ & 6.37 & 0. & 0. & 8.96$\times 10^{-2}$ \\
$b\bar{b}/\tau^+ \tau^- $ & 8.41 & -3.93$\times 10^{-2}$ & -1.86$\times 10^{-1}$ & 5.85$\times 10^{-2}$ & -2.46 & 2.57 & -4.59$\times 10^{-1}$ & 0. & 0. & -6.26$\times 10^{-2}$ \\
 $\tau^+\tau^-/cc$ & 2.58 & 9.29$\times 10^{-2}$ & 1.96$\times 10^{-1}$ & -3.11$\times 10^{-2}$ & 9.17 & 0. & 1.39 & -3.59 & 2.01 & 4.61$\times 10^{-2}$ \\
 $\gamma \gamma $/Z$\gamma $ & 1.57 & -1.19$\times 10^{-2}$ & -6.74 & 1.71 & 1.80$\times 10^{-2}$ & 1.88$\times 10^{-2}$ & 2.96 & 0. & 0. & -1.90 \\
$gg/Z\gamma $ & 3.31$\times 10^1$ & -1.47$\times 10^{-1}$ & -8.17 & -8.46$\times 10^{-1}$ & 2.49 & -6.12$\times 10^{-2}$ & 5.19 & 0. & 0. & -1.62 \\	  \hline \hline  
	\end{tabular}
	\caption{The reference value and expansion coefficients for ratios of Higgs boson decay branching fractions according to \eqx{eq:ratiobranchingfractiondefition}. The input parameters for this computation are from \tablex{tab:param}.  $VV^*$ partial decay widths are calculated by {\it Prophecy4f} in this table.}
	\label{tab:BRratio}
	\end{adjustwidth}
	\end{center}		
\endtable

\begin{center}
\begintable[t]
\centering
	\begin{tabular}{|c|c|c|c|c|c|}
		\hline\hline
 &  $P^\pm(\text{par.-add.})$ & $P^\pm(\text{par.-quad.})$ & $(P^+,\, P^-)(\mu)$ \\
	  \hline 
$\gamma \gamma /WW^*$ & 3.08 (0.85) & 2.98 (0.99) & (1.71,1.75) \\
$b\bar{b}/c\bar{c}$ & 22.27 (22.26) & 15.89 (15.89) & (0.62,0.41) \\
$\tau^+ \tau^- /\mu^+ \mu^- $ & 0.02 (0.02) & 0.02 (0.02) & (0.02,0.02) \\
$c\bar{c}/\mu^+ \mu^-$ & 20.58 (20.54) & 15.99 (15.99) & (0.46,0.34) \\
$WW^*/ZZ^*$ & 0.64 (0.25) & 0.52 (0.16) & (0.,0.) \\
$\gamma\gamma/ZZ^*$ & 3.61 (0.99) & 3.49 (0.88) & (1.71,1.75) \\
$b\bar{b}/ZZ^*$ & 9.32 (5.87) & 5.81 (3.72) & (0.31,0.02) \\
$\tau^+\tau^-/ZZ^*$ & 4.61 (1.20) & 4.55 (1.14) & (0.01,0.01) \\
$Z\gamma /ZZ^*$ & 1.41 (0.40) & 1.35 (0.34) & (0.73,0.71) \\
$b\bar{b}/\tau^+ \tau^- $ & 4.76 (4.71) & 3.53 (3.53) & (0.30,0.01) \\
$\tau^+ \tau^-/c\bar{c}$ & 20.60 (20.55) & 15.99 (15.99) & (0.33,0.44) \\
$\gamma \gamma /Z\gamma $ & 2.22 (0.61) & 2.15 (0.54) & (0.97,1.04) \\
$gg/Z\gamma$ & 4.25 (2.30) & 2.99 (1.61) & (0.79,2.94) \\
 	\hline \hline                       
	\end{tabular}
	\caption{This table gives the estimates for  theory error (percent relative uncertainty) of the ratio of branching fractions due to  parametric uncertainties and due to scale-dependent uncertainties from varying $m_H/2 \le \mu \le 2m_H$. Errors below 0.01\% are reported in the table as 0. For parametric uncertainties, we put an additional number in parentheses, which is the value it would have if the Higgs mass uncertainty were 0.1 GeV (instead of 0.4 GeV).}
	 \label{tab:BRratioUncertainties}
	\endtable
\end{center}

\section{Implications for Higgs Studies in the Standard Model and Beyond}

In this article we have done state-of-the art computations to detail the partial widths and branching fractions of the SM Higgs boson of $126\gev$. We have provided equations that Taylor expand the result about a set of input observables to show the shift in the partial width and branching fractions as a function of small deviations, including small deviations of the Higgs boson around $126\gev$. 

The purpose of computing Higgs boson properties is to enable precision comparisons of data with theory. Up to the present time, the experimental uncertainties for Higgs boson physics are much larger than the uncertainties of the theoretical computations. Furthermore, the SM predicted rates for Higgs observables are well within the bands of experimental measurements. 

Over time, however, the situation will change. It is hoped that experimental measurements will increase in precision so as to test any new physics contributions that might be influencing Higgs boson observables. The new physics contributions may be rather small and on the percent level~\cite{Gupta:2012mi}, and so it behooves us to come to an understanding of how precisely can one really test the SM Higgs boson couplings. Our analysis can be used to address that question as well.

For example, if the data at a later stage of the LHC, or ILC, or CLIC suggests that the branching fraction into $b$ quarks can be determined to better than 1\%, this does not mean that we are sensitive to new physics contributions of 1\% to $H\to b\bar b$. The reason can be seen from Tables \ref{tab:newBRerror} and~\ref{tab:BRfractionsUncertainties} that the SM uncertainty in computing $B(H\to b\bar b)$ is presently $3.7\%$ (sum of absolute values of all errors) and expected to not get better than $2.8\%$, with most of that coming from uncertainty of the bottom Yukawa coupling determination stemming from the uncertainty of the measured bottom quark pole mass, and the theory uncertainties encountered when extracting and connecting the two.
Thus, without reducing this error, any new physics contribution to the $b\bar b$ branching fraction that is not at least a factor of two or three larger than $2\%$  cannot be discerned. Thus, a deviation of at least $5\%$ is required of detectable new physics.

Some new physics ideas shift the $b$ quark Yukawa coupling away from the assumed SM value by virtue of the added contributions from induced finite $b$-quark mass corrections. An example of this is in high $\tan\beta$ supersymmetric theories. A relative shift in the $b$ quark Yukawa coupling $y_b\to y_b(1+\delta_b)$ translates into a shift of the branching fraction by $\delta B_b=2B_b(1-B_b)\delta_b\simeq \delta_b/2$. Thus, one would have to shift the bottom Yukawa coupling by more than $10\%$ to have any hope of discerning a non-SM signal. 

\begin{figure*}
\begin{center}
\includegraphics[width=0.75\columnwidth]{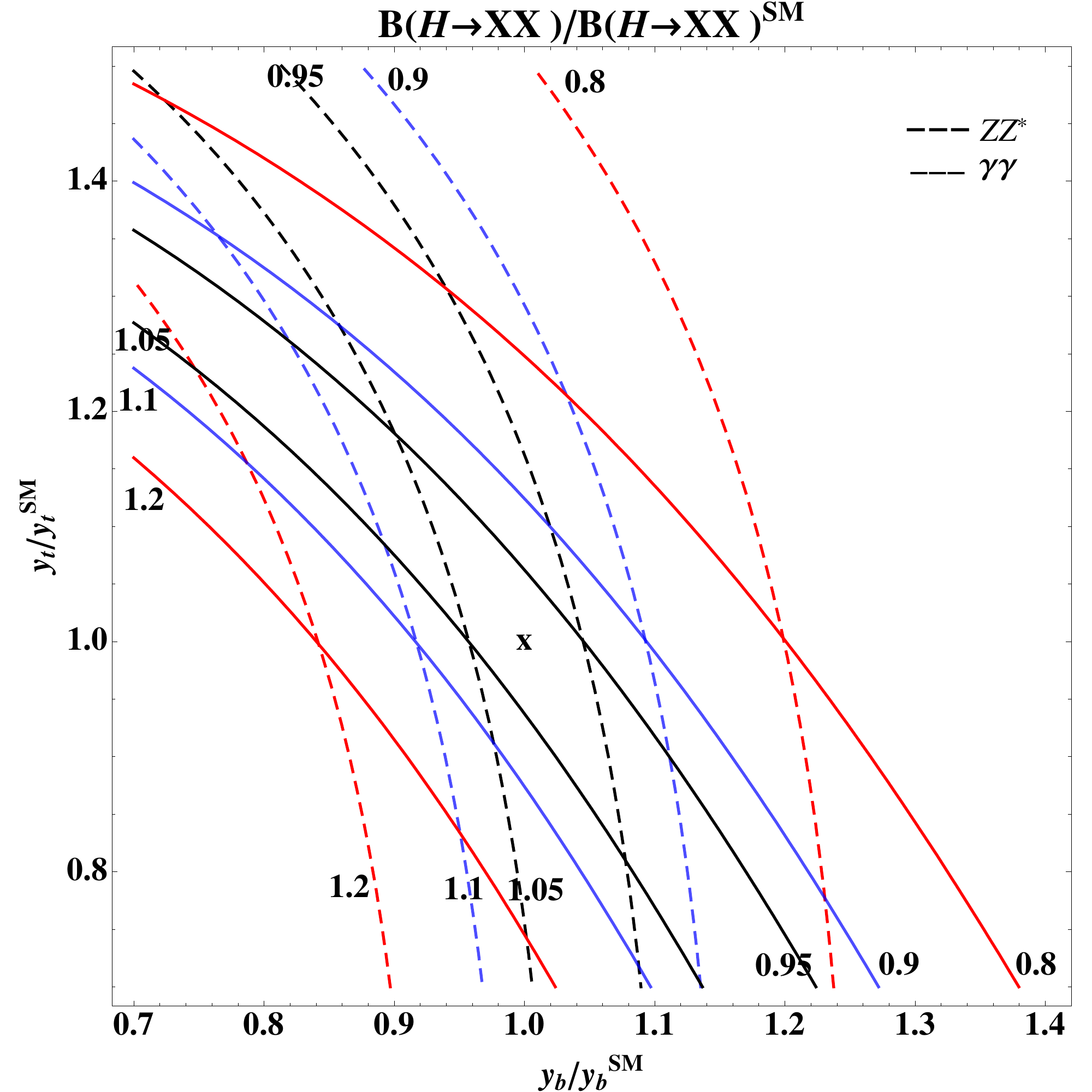}
\includegraphics[width=0.75\columnwidth]{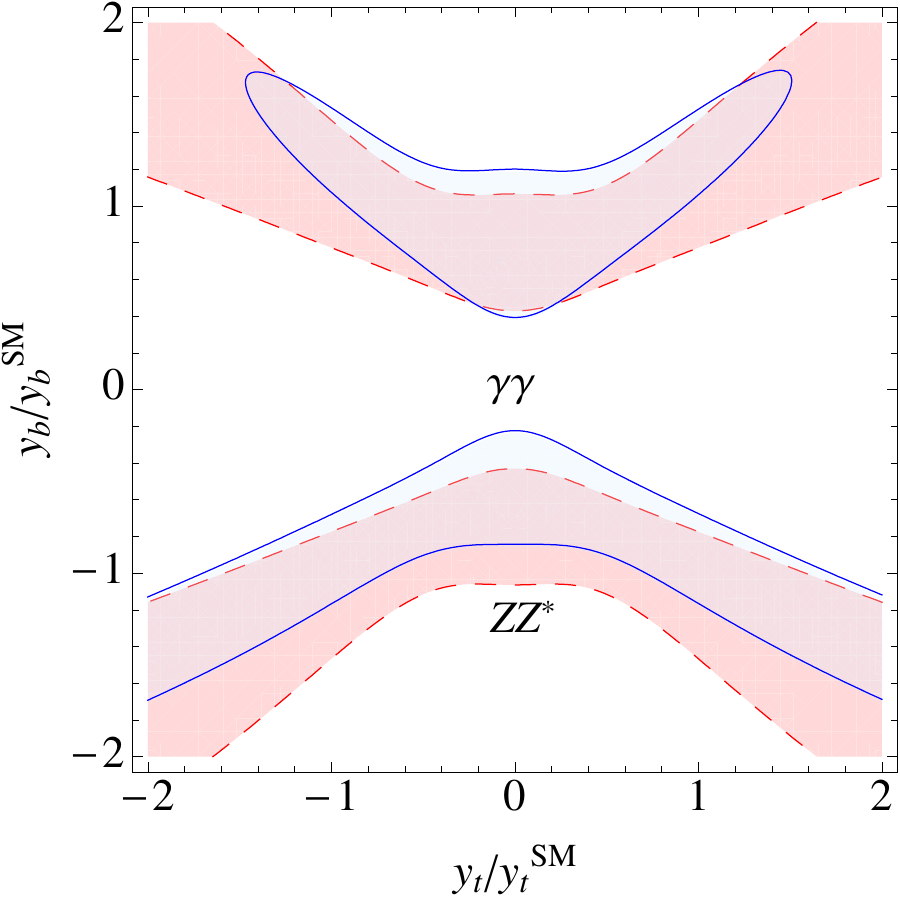}
\end{center}
\caption{{\bf Left Panel:} Contours of $B(H\to \gamma\gamma)/B(H\to \gamma\gamma)_{\rm SM}$ (solid lines) and $B(H\to ZZ)/B(H\to ZZ)_{SM}$ (dashed lines) in the $y_t-y_b$ plane. The SM position at $(1,1)$ is marked with an x. {\bf Right Panel:} The red shaded region is the $1\sigma$ allowed region for $y_t/y_t^{sm}$ and $y_b/y_b^{sm}$ given current data limits on $\sigma(H)\times B(H\to ZZ^*)$. The blue shaded region is the current $1\sigma$ allowed region from current data limits on $\sigma(H)\times B(H\to \gamma\gamma)$.}
\label{fig:BRoverBRSM1}
\end{figure*}

In the left panel of Fig.~\ref{fig:BRoverBRSM1}, we show contours of $B(H\to \gamma\gamma)/B(H\to \gamma\gamma)_{\rm SM}$ (solid lines) and $B(H\to ZZ)/B(H\to ZZ)_{SM}$ (dashed lines) in the $y_t-y_b$ plane, assuming that new physics only shifts the Yukawa couplings of the third generation fermions $t$ and $b$. The SM value for each is determined at the x position in the center of the figures where $y_f/y_f^{SM}=1$. At this point, the values are $B(H\to ZZ^*)=0.030$ and $B(H\to \gamma\gamma)=0.0027$.  The $1\sigma$ relative uncertainty of the SM calculation for  $\Delta B({ZZ^*})$ is about 4.0\% and for $\Delta B(\gamma\gamma)$ about 2.0\% (see \tablex{tab:BRfractionsUncertainties}).  These uncertainties cut large yet finite-width constraining areas in the plane of the left panel of Fig.~\ref{fig:BRoverBRSM1}. 
Furthermore, for correlated values of $\delta y_b$ and $\delta y_t$ shifts, there is no shift at all in these branching ratios. Nevertheless, the contours of constant $B(H\to \gamma\gamma)$ have different slope than the contours of $B(H\to ZZ)$, enabling determinations of $y_t$ and $y_b$ from a combination of precise measurements of these two observables alone.

In the right panel of Fig.~\ref{fig:BRoverBRSM1}, we demonstrate limits on $y_t$ and $y_b$ more directly from LHC data. The red shaded region is the $1\sigma$ allowed region for $y_t/y_t^{sm}$ and $y_b/y_b^{sm}$ given current data limits on $\sigma(H)\times B(H\to ZZ^*)$. The blue shaded region is the current $1\sigma$ allowed region from current data limits on $\sigma(H)\times B(H\to \gamma\gamma)$. The overlap region of these two shaded regions is the first estimate of where a global fit to the data suggests $y_t$ and $y_b$ must be. The $\gamma\gamma$ and $ZZ$ observables are the most powerful ones at present, and so it is appropriate to use them as illustration. 

As we can see, the data allows increasing $y_b$ as long as $y_t$ is increasing. This can be understood as the cancelation of two effects. When $y_b$ increases, the branching ratio to $bb$ increases, and therefore the branching fraction to $\gamma\gamma$ and $ZZ^*$ diminishes. However, if $y_t$ increases, then the production cross section $\sigma(gg\to H)$ increases, due to its primary contribution from a top quark loop diagram. It is these considerations that yield the shape of the allowed shaded regions in the right panel of Fig.~\ref{fig:BRoverBRSM1}. In the case of $\gamma\gamma$, an ever increasing positive $y_b$ and positive $y_t$ is not without bound because the production cross-section and branching fraction both increase. However, for positive $y_b$ and negative $y_t$, and vice versa, can be no bound since the production cross-section gains can be countered by the branching fraction loses, due to destructive interference, and vice versa.  In all cases, there is still room for quite sizable shifts in the top and bottom quark couplings to the Higgs boson while remaining consistent with the data. 


In conclusion, inspection of Tables~\ref{tab:BRfractionsUncertainties} and~\ref{tab:BRratioUncertainties} suggests that among most branching ratios, and among most ratios of branching ratios, the SM value cannot be determined theoretically to within better than a few percent.  When considering a future precision Higgs program at the LHC or another collider experiment beyond it, with hopes of getting measurements at the percent level or better to test new physics ideas, it will become necessary to confront the theory and input observable uncertainties that plague further improvement. We believe that the expansion technique presented in this paper is the most up-to-date presentation of partial width and branching ratio observable calculations in the SM, that it is ideal for investigating consequences of physics beyond the SM, and it most clearly shows the precise areas of improvement needed for SM calculations.

\section{Addendum: Results using $\overline{MS}$ $m_{b,c}$ inputs}

In this addendum, we replace our input parameters of bottom and charm quark pole mass by their $\overline{MS}$ masses. The new input parameters set is given in Table~\ref{tab:param-2}. 

The pole masses of Table~\ref{tab:param}, quoted from the Particle Data Group~\cite{Beringer:1900zz} and used in the previous results, were obtained from converting $\overline{MS}$ or $1S$ masses by two-loop conversion formulas. The analyses of the previous sections were essentially undoing these loop formulas to obtain $\overline{MS}$ masses, which has the effect of increasing the uncertainties in the partial widths due to the uncertainties in these masses compared to using $\overline{MS}$ masses to begin with. That is why we reproduce all of our results using the $\overline{MS}$ masses of $m_b$ and $m_c$.  To reduce the errors further more direct inputs from lattice computations~\cite{McNeile:2010ji} can help~\cite{Lepage:2014}.

Tables~\ref{tab:runmass-2}-\ref{tab:BRratioUncertainties-2} are reproductions of Tables~\ref{tab:runmass}-\ref{tab:BRratioUncertainties} using the $\overline{MS}$ inputs of Table~\ref{tab:param-2} rather than the pole mass inputs of Table~\ref{tab:param}. The figure captions are intended to be complete descriptions of what is calculated in each of these subsequent tables. The corresponding earlier pole mass inputs tables are referenced in the captions as well to make comparisons for the reader more convenient. As can be seen from the new entries of the tables, some of the dependences on the parameters have changed. For example, the dependence on $\alpha_s$ in the $b\bar b$ and $c\bar c$ channels is weaker due to change from pole to $\overline{MS}$ mass inputs. This in turn also implies a weaker dependence on $M_Z$, whose impact is from the scale where the $\alpha_s(M_Z)$ input is set. Overall, the uncertainties in the partial widths and branching fractions are reduced somewhat, especially for quark final states, by using $\overline{MS}$ quark mass inputs, as expected.

\begintable[t]
\centering
\begin{tabular}{ |c c | c c |}
\hline \hline
$m_H$ & 125.7(4)            & pole mass $m_t$           & 173.07(89)  \\
$\overline{\rm{MS}}$ mass $m_c$ & 1.275(25)   & $\overline{\rm{MS}}$ mass $m_b$ & 4.18(3)  \\
pole mass $m_\tau$ & 1.77682(16)         &    $\as (M_Z)$ & 0.1184(7) \\ 
$\alpha (M_Z)$  & ${1/128.96(2)}$     &   $\ahad$         & 0.0275(1) \\
\hline \hline
\end{tabular}
\caption{Reference values for the input observables, see \eqx{eq:parameters}, chosen for computation of the widths and branching ratios of the Higgs boson. Units are in GeV for the masses. All the reference values except for $m_H$~\cite{CMSHiggsMass} and $\alpha (M_Z)$ (or $\ahad$) are given by~\cite{Beringer:1900zz}. $\as (M_Z)$ is taken to be the world average value. As explained in the text, specifying $\alpha(M_Z)$ and $\ahad$, from the Winter 2012 plots of the LEPEWWG~\cite{LEPEWWGalpha}, in this table is redundant but done for convenience of the reader. These results were computed using $\overline{MS}$ $m_b$ and $m_c$ inputs (see Table~\ref{tab:param-2}) rather than their pole mass inputs (see Table~\ref{tab:param}). Compare results with the pole mass input results of Table~\ref{tab:param}.}
\label{tab:param-2}
\endtable

	\begintable[t]
		\centering
	\begin{tabular}{| c | c | c | }
	\hline \hline 
	quark & at $\mu=m_H\, (m_H/2,\,2m_H)$ & $P_{m}(\Delta m)$ \\
	\hline 
	$m_c (\mu)$ & 0.638 (0.675, 0.603) GeV & 2.62\% \\ 
	\hline	
	$m_b (\mu)$ & 2.79 (2.96, 2.64) GeV & 0.85\% \\ 
	\hline
	$m_t (\mu)$ & 166.5 (176.8,157.8) GeV & 0.56\% \\ 
	\hline \hline
	\end{tabular}
	\caption{Running $\overline{\rm MS}$ masses for the heavy quarks at 3-loops at the scale $\mu=m_H$, $m_H/2$ and $2m_H$ from program RunDec~\cite{Chetyrkin:2000yt}, which is used for the Higgs decaying into quarks. Pole-mass inputs are taken from \tablex{tab:param}. The parametric uncertainty on the running mass at $\mu=m_H$ from $1\sigma$ uncertainty ($\sigma_m$) in the pole mass is defined to be $P_m(\Delta m)=\{ m_+(m_H)+m_-(m_H)\}/\{2m(m_H)\}$, where $m_\pm(m_H)$ is computed using $m_{\rm pole}=m_{\rm ref}\pm \sigma_m$. The scale dependence of the running mass is cancelled in higher order loop calculations, as can be seen later for scale-dependence uncertainties. These results were computed using $\overline{MS}$ $m_b$ and $m_c$ inputs (see Table~\ref{tab:param-2}) rather than their pole mass inputs (see Table~\ref{tab:param}). Compare results with the pole mass input results of Table~\ref{tab:runmass}.}
	\label{tab:runmass-2}
	\endtable

\begintable[t]
\begin{center}
                \footnotesize
	\begin{tabular}{|c|c|c|c|c|c|c|c|c|c|c|c|}
		\hline\hline
 &   $\Gamma_X^{\rm(Ref)}/{\rm GeV}$ & $a_{m_t,X}$ &  $a_{m_H,X}$ &  $a_{\alpha ( M_Z),X}$ & $a_{\as ( M_Z),X}$ &  $a_{m_b,X}$ &  $a_{M_Z,X}$ &  $a_{m_c,X} $ & $a_{m_\tau,X}$ & $a_{G_F,X}$\\
	  \hline 
$\textrm{total}$ & 4.17$\times 10^{-3}$ & -3.3$\times 10^{-2}$ & 4.34 & 8.35$\times 10^{-1}$ & -5.05$\times 10^{-1}$ & 1.32 & -3.21 & 7.80$\times 10^{-2}$ & 1.24$\times 10^{-1}$ & 8.49$\times 10^{-1}$ \\
$gg$ & 3.61$\times 10^{-4}$ & -1.62$\times 10^{-1}$ & 2.89 & 0. & 2.48 & -6.51$\times 10^{-2}$ & 3.76$\times 10^{-1}$ & 0. & 0. & 1.00 \\
$\gamma\gamma$ & 1.08$\times 10^{-5}$ & -2.69$\times 10^{-2}$ & 4.32 & 2.56 & 1.80$\times 10^{-2}$ & 8.29$\times 10^{-3}$ & -1.86 & 0. & 0. & 7.24$\times 10^{-1}$ \\
$b\bar{b}$ & 2.35$\times 10^{-3}$ & 8.07$\times 10^{-3}$ & 8.09$\times 10^{-1}$ & 3.76$\times 10^{-2}$ & -1.12 & 2.36 & -2.72$\times 10^{-1}$ & 0. & 0. & 9.53$\times 10^{-1}$ \\
 $c\bar{c}$ & 1.22$\times 10^{-4}$ & -4.52$\times 10^{-2}$ & 7.99$\times 10^{-1}$ & 1.02$\times 10^{-2}$ & -3.10 & 0. & -4.89$\times 10^{-1}$ & 2.67 & 0. & 9.70$\times 10^{-1}$ \\
$\tau^+\tau^-$ & 2.58$\times 10^{-4}$ & 4.71$\times 10^{-2}$ & 9.95$\times 10^{-1}$ & -2.09$\times 10^{-2}$ & -2.14$\times 10^{-3}$ & 0. & -1.61$\times 10^{-2}$ & 0. & 2.01 & 1.02 \\
$WW^*$ & 9.43$\times 10^{-4}$ & -1.13$\times 10^{-1}$ & 1.37$\times 10^1$ & 3.66 & 9.04$\times 10^{-3}$ & 0. & -1.21$\times 10^1$ & 0. & 0. & 2.49$\times 10^{-1}$ \\
$ZZ^*$ & 1.17$\times 10^{-4}$ & 2.27$\times 10^{-2}$ & 1.53$\times 10^1$ & -7.37$\times 10^{-1}$ & -1.82$\times 10^{-3}$ & 0. & -1.12$\times 10^1$ & 0. & 0. & 2.53 \\
$ Z\gamma$ & 6.89$\times 10^{-6}$ & -1.52$\times 10^{-2}$ & 1.11$\times 10^1$ & 8.45$\times 10^{-1}$ & 0. & -7.93$\times 10^{-3}$ & -4.82 & 0. & 0. & 2.62 \\
 $\mu^+\mu^-$ & 8.93$\times 10^{-7}$ & 4.82$\times 10^{-2}$ & 9.92$\times 10^{-1}$ & -4.31$\times 10^{-2}$ & -2.19$\times 10^{-3}$ & 0. & -1.62$\times 10^{-2}$ & 0. & 0. & 1.02 \\

	\hline \hline                       
	\end{tabular}
	\caption{Reference values for the partial widths at the central values of the parameters given in 
	         \tablex{tab:param} along with values for $a_{\tau_i,X}$ as
  			 defined by \eqx{eq:partialwidthdefition}. $VV^*$ partial decay widths are calculated by {\it Prophecy4f}. These results were computed using $\overline{MS}$ $m_b$ and $m_c$ inputs (see Table~\ref{tab:param-2}) rather than their pole mass inputs (see Table~\ref{tab:param}). Compare results with the pole mass input results of Table~\ref{tab:analresults1}.}.
	 \label{tab:analresults1-2}
	 \end{center}
	\endtable

\begintable[t]
\begin{center}
	\begin{tabular}{|c|c|c|c|}
		\hline\hline
 &  $P_\Gamma^\pm(\text{par.add.})$ & $P_\Gamma^\pm(\text{par.quad.})$ & $(P_\Gamma^+,\, P_\Gamma^-)(\mu)$ \\
	  \hline 
\textrm{total} & 2.82 (1.79) & 1.71 (1.07) & (0.08,0.10) \\
$gg$ & 2.52 (1.83) & 1.74 (1.49) & (0.05,0.03) \\
$\gamma\gamma$  & 1.45 (0.42) & 1.38 (0.35) & (1.31,0.60) \\
 $b\bar{b}$ & 2.62 (2.43) & 1.84 (1.82) & (0.29,0.01) \\
 $c\bar{c}$ & 7.34 (7.15) & 5.55 (5.54) & (0.45,0.35) \\
$\tau^+\tau^-$   & 0.36 (0.12) & 0.32 (0.08) & (0.01,0.01) \\
$WW^*$ & 4.41 (1.17) & 4.97 (1.25) & (0.25,0.31) \\
$ZZ^*$ & 4.90 (1.25) & 4.42 (1.11) & (0.,0.) \\
$Z\gamma$ & 3.56 (0.92) & 3.52 (0.88) & (0.56,0.23) \\
$\mu^+\mu^-$  & 0.34 (0.11) & 0.32 (0.08) & (0.03,0.03) \\
  	\hline \hline                       
	\end{tabular}
	\caption{This table gives the estimates for percent relative uncertainty on the partial widths from parametric and scale-dependence uncertainties. Parametric uncertainties arise from incomplete knowledge of the input observables for the calculation (i.e., errors on $m_c$, $\alpha_s$, etc.). For parametric uncertainties, we put an additional number in parentheses, which is the value it would have if the Higgs mass uncertainty were 0.1 GeV (instead of 0.4 GeV).
	Scale-dependence uncertainties are indicative of not knowing the higher order terms in a perturbative expansion of the observable. These uncertainties are estimated by varying $\mu$ from $m_H/2$ to $2m_H$.  More details on the precise meaning of the entries of this table are found in the text of sec.~\ref{sec:partial}. Errors below $0.01\%$ are represented in this table as 0. These results were computed using $\overline{MS}$ $m_b$ and $m_c$ inputs (see Table~\ref{tab:param-2}) rather than their pole mass inputs (see Table~\ref{tab:param}). Compare results with the pole mass input results of Table~\ref{tab:theoreticaluncertainties}.}
	 \label{tab:theoreticaluncertainties-2}
\end{center}
	\endtable

 \begintable[t]  %
 \begin{center}
	\footnotesize
	\begin{tabular}{|c|c|c|c|c|c|c|c|c|c|c|}
		\hline \hline
&  $\rm{B}(X)^{(\rm{Ref})}$ &  $b_{m_t}$ &  $b_{m_H}$ &  $b_{\alpha( M_Z)}$ &  $b_{\as( M_Z)}$ &  $b_{m_b}$ &  $b_{M_Z}$ &  $b_{m_c} $ & $b_{m_\tau}$ & $b_{G_F}$ \\
	   \hline 	
$gg$ & 8.68$\times 10^{-2}$ & -1.29$\times 10^{-1}$ & -1.46 & -8.35$\times 10^{-1}$ & 2.99 & -1.39 & 3.58 & -7.8$\times 10^{-2}$ & -1.24$\times 10^{-1}$ & 1.51$\times 10^{-1}$ \\
$\gamma\gamma$ & 2.58$\times 10^{-3}$ & 6.09$\times 10^{-3}$ & -2.12$\times 10^{-2}$ & 1.73 & 5.23$\times 10^{-1}$ & -1.32 & 1.35 & -7.80$\times 10^{-2}$ & -1.24$\times 10^{-1}$ & -1.25$\times 10^{-1}$ \\
$b\bar{b}$ & 5.63$\times 10^{-1}$ & 4.10$\times 10^{-2}$ & -3.54 & -7.98$\times 10^{-1}$ & -6.16$\times 10^{-1}$ & 1.04 & 2.93 & -7.8$\times 10^{-2}$ & -1.24$\times 10^{-1}$ & 1.04$\times 10^{-1}$ \\
$c\bar{c}$ & 2.92$\times 10^{-2}$ & -1.23$\times 10^{-2}$ & -3.55 & -8.25$\times 10^{-1}$ & -2.59 & -1.32 & 2.72 & 2.59 & -1.24$\times 10^{-1}$ & 1.21$\times 10^{-1}$ \\
$\tau^+\tau^-$ & 6.18$\times 10^{-2}$ & 8.01$\times 10^{-2}$ & -3.35 & -8.56$\times 10^{-1}$ & 5.03$\times 10^{-1}$ & -1.32 & 3.19 & -7.80$\times 10^{-2}$ & 1.88 & 1.67$\times 10^{-1}$ \\
$WW^*$ & 2.26$\times 10^{-1}$ & -7.99$\times 10^{-2}$ & 9.32 & 2.82 & 5.14$\times 10^{-1}$ & -1.32 & -8.91 & -7.8$\times 10^{-2}$ & -1.24$\times 10^{-1}$ & -5.99$\times 10^{-1}$ \\
$ZZ^*$ & 2.81$\times 10^{-2}$ & 5.57$\times 10^{-2}$ & 1.10$\times 10^1$ & -1.57 & 5.03$\times 10^{-1}$ & -1.32 & -7.98 & -7.80$\times 10^{-2}$ & -1.24$\times 10^{-1}$ & 1.68 \\
$Z\gamma$ & 1.65$\times 10^{-3}$ & 1.78$\times 10^{-2}$ & 6.71 & 9.89$\times 10^{-3}$ & 5.05$\times 10^{-1}$ & -1.33 & -1.61 & -7.80$\times 10^{-2}$ & -1.24$\times 10^{-1}$ & 1.77 \\
$\mu^+\mu^-$ & 2.14$\times 10^{-4}$ & 8.11$\times 10^{-2}$ & -3.35 & -8.79$\times 10^{-1}$ & 5.03$\times 10^{-1}$ & -1.32 & 3.19 & -7.80$\times 10^{-2}$ & -1.24$\times 10^{-1}$ & 1.67$\times 10^{-1}$ \\ 	  \hline \hline  
	\end{tabular}
	\caption{The reference value and expansion coefficients for Higgs boson decay branching fractions according to \eqx{eq:branchingfractiondefition}. The input parameters for this computation are from \tablex{tab:param}.
	 $VV^*$ partial decay widths are calculated by {\it Prophecy4f}. These results were computed using $\overline{MS}$ $m_b$ and $m_c$ inputs (see Table~\ref{tab:param-2}) rather than their pole mass inputs (see Table~\ref{tab:param}). Compare results with the pole mass input results of Table~\ref{tab:BRfractions1}.}
	\label{tab:BRfractions1-2}
\end{center}
\endtable

 \begintable[t]  %
 \begin{center}
	\footnotesize
	\begin{tabular}{|c|c|c|c|c|c|c|c|c|c|}
		\hline \hline
& $\Delta_{m_t}$ & $\Delta_{m_H}$ & $\Delta_{\alpha(M_Z)}$ & $\Delta_{\as(M_Z)}$ & $\Delta_{m_b}$ & $\Delta_{M_Z}$ & $\Delta_{m_c}$ & $\Delta_{m_\tau}$ & $\Delta_{G_F}$ \\		
	   \hline 	
$gg$ & 0.07 & 0.46 (0.12) & 0.01 & 1.77 & 1.00 & 0.01 & 0.15 & - & - \\	   
$\gamma\gamma$ & - & 0.01 ( - ) & 0.03 & 0.31 & 0.94 & - & 0.15 & - & - \\
$b\bar{b}$ & 0.02 & 1.13 (0.28) & 0.01 & 0.36 & 0.74 & 0.01 & 0.15 & - & - \\
$c\bar{c}$ & 0.01 & 1.13 (0.28) & 0.01 & 1.53 & 0.95 & 0.01 & 5.08 & - & - \\
$\tau^+\tau^-$ & 0.04 & 1.07 (0.27) & 0.01 & 0.30 & 0.95 & 0.01 & 0.15 & 0.02 & - \\
$WW^*$ & 0.04 & 2.97 (0.74) & 0.04 & 0.30 & 0.95 &0.02 & 0.15 & - & - \\
$ZZ^*$ & 0.03 & 3.48 (0.87) & 0.02 & 0.30 & 0.95 & 0.02 & 0.15 & - & - \\
$Z\gamma$ & 0.01 & 2.14 (0.53) & - & 0.30 & 0.96 & - & 0.15 & - & - \\
$\mu^+\mu^-$ & 0.04 & 1.07 (0.27) & 0.01 & 0.30 & 0.95 & 0.01 & 0.15 & - & - \\
 	  \hline \hline  
	\end{tabular}
	\caption{Table of percentage uncertainties of branching fractions due to uncertainties in each of the input observables, as calculated by eq.~\ref{eq:percentuncertainty}. The input parameters for this computation are from \tablex{tab:param}. In addition we also compute the branching ratio uncertainties due to $\Delta m_h = 0.1$ GeV, the expected uncertainty after LHC run. These values are in parenthesis in the $\Delta_{m_H}$ column. Percentages less than $0.1\%$ are listed as $-$.
These results were computed using $\overline{MS}$ $m_b$ and $m_c$ inputs (see Table~\ref{tab:param-2}) rather than their pole mass inputs (see Table~\ref{tab:param}). Compare results with the pole mass input results of Table~\ref{tab:newBRerror}.}	\label{tab:newBRerror-2}
\end{center}
\endtable

 
\begintable[t]
 \begin{center}
	\begin{tabular}{|c|c|c|c|}
		\hline\hline
 &  $P_\text{BR}^\pm(\text{par.-add.})$ & $P_\text{BR}^\pm(\text{par.-quad.})$ & $(P_\text{BR}^+,\, P_\text{BR}^-)(\mu)$ \\
	  \hline 
$gg$ &  3.47 (3.12) &  2.09 (2.04) &  (0.03,1.38) \\
$\gamma\gamma$  &  1.45 (1.44) &  1.01 (1.01) &  (1.81,1.83) \\
$b\bar{b}$ &  2.43 (1.58) &  1.41 (0.89) &  (0.21,0.) \\
$c\bar{c}$ &  8.72 (7.87) &  5.51 (5.40) &  (0.54,0.44) \\
$\tau^+\tau^-$ &  2.55 (1.75) &  1.47 (1.04) &  (0.09,0.07) \\
$WW^*$ &  4.48 (2.26) &  3.13 (1.25) &  (0.10,0.08) \\
$ZZ^*$ &  4.96 (2.34) &  3.63 (1.33) &  (0.10,0.08) \\
$Z\gamma$ &  3.56 (1.96) &  2.36 (1.15) &  (0.83,0.80) \\
$\mu^+\mu^-$  &  2.53 (1.73) &  1.47 (1.04) &  (0.07,0.06) \\
 	\hline \hline                       
	\end{tabular}
	\caption{This table gives the estimates for theory error (percent relative uncertainty) of the branching fractions due to  parametric uncertainties and due to scale-dependent uncertainties from varying $m_H/2 \le \mu \le 2m_H$. Errors below 0.01\% are reported in the table as 0. For parametric uncertainties, we put an additional number in parentheses, which is the value it would have if the Higgs mass uncertainty were 0.1 GeV (instead of 0.4 GeV). These results were computed using $\overline{MS}$ $m_b$ and $m_c$ inputs (see Table~\ref{tab:param-2}) rather than their pole mass inputs (see Table~\ref{tab:param}). Compare results with the pole mass input results of Table~\ref{tab:BRfractionsUncertainties}.}
	 \label{tab:BRfractionsUncertainties-2}
\end{center}
	\endtable
	
\begintable[t]
\begin{center}
		\footnotesize
\begin{adjustwidth}{-0.0cm}{}	
	\begin{tabular}{|c|c|c|c|c|c|c|c|c|c|c|c|c|}
		\hline\hline
{} & $B(X)/B(Y)_{Ref}$ & $r_{m_t}$ & $r_{m_H}$ & $r_{\alpha(M_Z)}$ & $r_{\as(M_Z)}$ & $r_{m_b}$ & $r_{M_Z}$ & $r_{m_c}$ & $r_{m_\tau}$  & $r_{G_F}$ \\
	   \hline 	
$\gamma \gamma /WW^*$ & 1.14$\times 10^{-2}$ & 8.60$\times 10^{-2}$ & -9.35 & -1.10 & 8.99$\times 10^{-3}$ & 8.29$\times 10^{-3}$ & 1.03$\times 10^1$ & 0. & 0. & 4.75$\times 10^{-1}$ \\
$b\bar{b}/c\bar{c}$ & 1.93$\times 10^1$ & 5.33$\times 10^{-2}$ & 1.01$\times 10^{-2}$ & 2.74$\times 10^{-2}$ & 1.98 & 2.36 & 2.17$\times 10^{-1}$ & -2.67 & 0. & -1.67$\times 10^{-2}$ \\
$\tau^+ \tau^- /\mu^+ \mu^- $ & 2.89$\times 10^2$ & -1.02$\times 10^{-3}$ & 2.55$\times 10^{-3}$ & 2.22$\times 10^{-2}$ & 4.63$\times 10^{-5}$ & 0. & 1.09$\times 10^{-4}$ & 0. & 2.01 & -3.36$\times 10^{-4}$ \\
$c\bar{c}/\mu^+ \mu^-$ & 1.36$\times 10^2$ & -9.34$\times 10^{-2}$ & -1.93$\times 10^{-1}$ & 5.33$\times 10^{-2}$ & -3.10 & 0. & -4.73$\times 10^{-1}$ & 2.67 & 0. & -4.62$\times 10^{-2}$ \\
 $WW^*/ZZ^*$ & 8.05 & -1.36$\times 10^{-1}$ & -1.63 & 4.40 & 1.09$\times 10^{-2}$ & 0. & -9.38$\times 10^{-1}$ & 0. & 0. & -2.28 \\
$\gamma\gamma/ZZ^*$ & 9.19$\times 10^{-2}$ & -4.96$\times 10^{-2}$ & -1.10$\times 10^1$ & 3.30 & 1.99$\times 10^{-2}$ & 8.29$\times 10^{-3}$ & 9.33 & 0. & 0. & -1.81 \\
$b\bar{b}/ZZ^*$ & 2.00$\times 10^1$ & -1.47$\times 10^{-2}$ & -1.45$\times 10^1$ & 7.74$\times 10^{-1}$ & -1.12 & 2.36 & 1.09$\times 10^1$ & 0. & 0. & -1.58 \\
$\tau^+\tau^-/ZZ^*$ & 2.20 & 2.44$\times 10^{-2}$ & -1.43$\times 10^1$ & 7.16$\times 10^{-1}$ & -3.19$\times 10^{-4}$ & 0. & 1.12$\times 10^1$ & 0. & 2.01 & -1.52 \\
$Z\gamma /ZZ^*$ & 5.88$\times 10^{-2}$ & -3.79$\times 10^{-2}$ & -4.24 & 1.58 & 1.82$\times 10^{-3}$ & -7.93$\times 10^{-3}$ & 6.37 & 0. & 0. & 9.$\times 10^{-2}$ \\
$b\bar{b}/\tau^+ \tau^- $ & 9.11 & -3.91$\times 10^{-2}$ & -1.85$\times 10^{-1}$ & 5.85$\times 10^{-2}$ & -1.12 & 2.36 & -2.56$\times 10^{-1}$ & 0. & 0. & -6.25$\times 10^{-2}$ \\
 $\tau^+\tau^-/cc$ & 2.12 & 9.24$\times 10^{-2}$ & 1.96$\times 10^{-1}$ & -3.11$\times 10^{-2}$ & 3.10 & 0. & 4.73$\times 10^{-1}$ & -2.67 & 2.01 & 4.58$\times 10^{-2}$ \\
 $\gamma \gamma $/Z$\gamma $ & 1.56 & -1.17$\times 10^{-2}$ & -6.74 & 1.72 & 1.80$\times 10^{-2}$ & 1.62$\times 10^{-2}$ & 2.96 & 0. & 0. & -1.90 \\
$gg/Z\gamma $ & 3.31$\times 10^1$ & -1.47$\times 10^{-1}$ & -8.17 & -8.45$\times 10^{-1}$ & 2.48 & -5.72$\times 10^{-2}$ & 5.19 & 0. & 0. & -1.62 \\
	  \hline \hline  
	\end{tabular}
	\caption{The reference value and expansion coefficients for ratios of Higgs boson decay branching fractions according to \eqx{eq:ratiobranchingfractiondefition}. The input parameters for this computation are from \tablex{tab:param}.  $VV^*$ partial decay widths are calculated by {\it Prophecy4f} in this table. These results were computed using $\overline{MS}$ $m_b$ and $m_c$ inputs (see Table~\ref{tab:param-2}) rather than their pole mass inputs (see Table~\ref{tab:param}). Compare results with the pole mass input results of Table~\ref{tab:BRratio}.}
	\label{tab:BRratio-2}
	\end{adjustwidth}
	\end{center}		
\endtable

\begin{center}
\begintable[t]
\centering
	\begin{tabular}{|c|c|c|c|c|c|}
		\hline\hline
 &  $P^\pm(\text{par.-add.})$ & $P^\pm(\text{par.-quad.})$ & $(P^+,\, P^-)(\mu)$ \\
	  \hline 
$\gamma \gamma /WW^*$  &  3.71 (1.48)  &  3.04 (0.99)  &  (1.71,1.75)  \\
$b\bar{b}/c\bar{c}$  &  8.13 (8.12)  &  5.62 (5.62)  &  (0.65,0.42)  \\
$\tau^+ \tau^- /\mu^+ \mu^- $  &  0.02 (0.02)  &  0.02 (0.02)  &  (0.02,0.02)  \\
$c\bar{c}/\mu^+ \mu^-$  &  7.17 (7.13)  &  5.54 (5.54)  &  (0.47,0.38)  \\
$WW^*/ZZ^*$  &  0.66 (0.28)  &  0.53 (0.16)  &  (0.,0.)  \\
$\gamma\gamma/ZZ^*$  &  3.61 (0.99)  &  3.49 (0.88)  &  (1.71,1.75)  \\
$b\bar{b}/ZZ^*$  &  7.01 (3.55)  &  4.96 (2.15)  &  (0.29,0.01)  \\
$\tau^+\tau^-/ZZ^*$  &  4.62 (1.21)  &  4.55 (1.14)  &  (0.01,0.01)  \\
$Z\gamma /ZZ^*$  &  1.41 (0.40)  &  1.35 (0.34)  &  (0.73,0.71)  \\
$b\bar{b}/\tau^+ \tau^- $  &  2.44 (2.39)  &  1.82 (1.82)  &  (0.28,0.01)  \\
$\tau^+ \tau^-/c\bar{c}$  &  7.19 (7.14)  &  5.54 (5.54)  &  (0.36,0.45)  \\
$\gamma \gamma /Z\gamma $  &  2.21 (0.60)  &  2.14 (0.54)  &  (0.97,1.04)  \\
$gg/Z\gamma$  &  4.21 (2.26)  &  2.99 (1.61)  &  (0.99,3.11)  \\
 	\hline \hline                       
	\end{tabular}
	\caption{This table gives the estimates for  theory error (percent relative uncertainty) of the ratio of branching fractions due to  parametric uncertainties and due to scale-dependent uncertainties from varying $m_H/2 \le \mu \le 2m_H$. Errors below 0.01\% are reported in the table as 0. For parametric uncertainties, we put an additional number in parentheses, which is the value it would have if the Higgs mass uncertainty were 0.1 GeV (instead of 0.4 GeV). These results were computed using $\overline{MS}$ $m_b$ and $m_c$ inputs (see Table~\ref{tab:param-2}) rather than their pole mass inputs (see Table~\ref{tab:param}). Compare results with the pole mass input results of Table~\ref{tab:BRratioUncertainties}.}
	 \label{tab:BRratioUncertainties-2}
	\endtable
\end{center}

\bigskip \noindent 
{\it Acknowledgments:} 
We thank K. Chetyrkin for helpful communications. We thank A. Mueck, S. Heinemeyer, I. Puljak, and D. Rebuzzi for drawing our attention to corrections to our original $\alpha_s$ dependencies. We wish also to thank M.~Peskin for bringing to our attention the improvements in uncertainty when using  $\overline{MS}$ masses directly as inputs (see addendum for these results).
SL is supported by the National Research Foundation of Korea (NRF) grant funded by the Korea government (MEST) N01120547. LA was supported by the P2IO Labex.
SP have been supported by National Science Centre under research grants DEC-2011/01/M/ST2/02466, DEC-2012/04/A/ST2/00099, DEC-2012/05/B/ST2/02597. 
LA and SL would also like to thank the CERN theory group for their hospitality, where part of this work has been done.
This work was supported by the Supercomputing Center/Korea Institute of Science and Technology Information with supercomputing resources including technical support pd0326.

\section*{Appendix A}
In this appendix, we give the scale dependence to scalar correlator, $R_S(\mu^2 ,m_h^2)$ at  ${\cal{O}}(\as^4)$. This contributes to the decay of Higgs into heavy quarks in the following form

\begin{equation}
\Gamma_{H\to Q\bar{Q}} (\mu^2) = \bar{\sigma_0}  m^2_Q (\mu^2) R_S \left(\frac{\mu^2}{m_h^2}\right),
\end{equation}
where $\bar{\sigma_0}$ is lowest-order cross-section without the outgoing quark masses.
The scale invariance of $\Gamma_{H \to Q \bar{Q}} (\mu^2), $ together with the results for $R_S$ at  $\mu=m_h$ to ${\cal{O}}(\as^4)$ given in \cite{Baikov:2005rw},
leads to the following additional $\as^4 $ contributions to $R_S(\mu^2,m_h^2)$ previously not found in literature. 

\bea
R_S(\mu^2,m_h^2) &=& 
3 \left\{1 + \left( \frac{\as(\mu)}{\pi}\right) \left( 5.666 + 2 \log \left(\frac{\mu^2}{m_h^2}\right) \right)  + \right. \nonumber \\ 
&&\hspace{-3cm} \left. \left( \frac{\as(\mu)}{\pi}\right)^2 \left(29.1467 + 29.222 \log \left( \frac{\mu^2}{m_h^2}\right) + 
    3.917 \log^2 \left(\frac{\mu^2}{m_h^2}\right) \right) + \right. \nonumber \\
 &&\hspace{-3cm} \left. \left( \frac{\as(\mu)}{\pi}\right)^3 \left(41.758 + 185.295 \log \left(\frac{\mu^2}{m_h^2}\right)+ 90.545 \log^2 \left(\frac{\mu^2}{m_h^2}\right) \right. \right. \nonumber \\ 
&& \hspace{-1.5cm} \left. \left.  +7.616\log^3 \left(\frac{\mu^2}{m_h^2}\right)^3\right)  +\right. \nonumber \\
    &&\hspace{-3cm} \left. \left( \frac{\as(\mu)}{\pi}\right)^4 \left(-825.7 + 443.937   \log \left(\frac{\mu^2}{m_h^2}\right) + 721.581 \log^2 \left(\frac{\mu^2}{m_h^2}\right) \right.\right. \nonumber \\
    && \hspace{-1.5cm} \left. \left.  +  238.608 \log^3 \left(\frac{\mu^2}{m_h^2}\right) + 14.755 \log^4 \left(\frac{\mu^2}{m_h^2}\right) \right)  \right\}
\eea

The coefficients are given at $N_c=3$, $n_f=5$ since these are where the standard  model contributes and where the values at $\mu=m_h$ are reported. These contribute to an ${\cal{O}}(0.01\%)$ scale uncertainty to the width results.

\end{document}